\documentclass[10pt,conference]{IEEEtran}

\usepackage{times}
\usepackage{graphicx}
\usepackage{booktabs}
\usepackage{multirow}
\usepackage{hyperref}
\usepackage{amsmath}
\usepackage{enumitem}
\usepackage{longtable}
\usepackage{caption}
\usepackage{subcaption}
\usepackage{xcolor}
\usepackage{listings}
\usepackage{url}
\usepackage[none]{hyphenat}   % disable all automatic hyphenation
\usepackage{multicol}
\usepackage{float}
\usepackage{tabularx}
\usepackage{cuted}

\usepackage{fancyhdr}
\title{AI Bill of Materials and Beyond: Systematizing Security Assurance through the AI Risk Scanning (AIRS) Framework}

\author{
\IEEEauthorblockN{Samuel Nathanson, Alexander Lee, Catherine Chen Kieffer}
\IEEEauthorblockN{Jared Junkin, Jessica Ye, Amir Saeed}
\IEEEauthorblockN{Melanie Lockhart, Russ Fink,  Lanier Watkins, Elisha Peterson}
\IEEEauthorblockA{Johns Hopkins University Applied Physics Laboratory (APL)\\
\texttt{Corresponding Author: Alexander.Lee@jhuapl.edu}}
}

\begin{document}
\maketitle
\thispagestyle{fancy}

\begin{abstract}
Assurance for artificial intelligence (AI) systems remains fragmented across software supply-chain security, adversarial machine learning, and governance documentation. Existing transparency mechanisms---including Model Cards, Datasheets, and Software Bills of Materials (SBOMs)---advance provenance reporting but rarely provide verifiable, machine-readable evidence of model security. This paper introduces the \emph{AI Risk Scanning (AIRS)} Framework, a threat-model-based, evidence-generating framework designed to operationalize AI assurance. The AIRS Framework evolved through three progressive pilot studies---\emph{Smurf} (AIBOM schema design), \emph{OPAL} (operational validation), and \emph{Pilot~C} (AIRS)---that collectively reframed AI documentation from descriptive disclosure toward measurable, evidence-bound verification. The framework aligns its assurance fields to the MITRE~ATLAS adversarial ML taxonomy and automatically produces structured artifacts capturing model integrity, packaging and serialization safety, structural adapters, and runtime behaviors. Currently, the AIRS Framework is scoped to provide model-level assurances for LLMs, but the framework could be expanded to include other modalities and cover system-level threats (e.g. application-layer abuses, tool-calling). A proof-of-concept on a quantized \texttt{GPT-OSS-20B} model demonstrates enforcement of safe loader policies, per-shard hash verification, and contamination and backdoor probes executed under controlled runtime conditions. Comparative analysis with SBOM standards of SPDX~3.0 and CycloneDX~1.6 reveals alignment on identity and evaluation metadata, but identifies critical gaps in representing AI-specific assurance fields. The AIRS Framework thus extends SBOM practice to the AI domain by coupling threat modeling with automated, auditable evidence generation, providing a principled foundation for standardized, trustworthy, and machine-verifiable AI risk documentation.
\end{abstract}


\section{Introduction}
\label{sec:intro}

Assurance and security documentation for artificial intelligence (AI) systems remain fragmented across the disciplines of software supply-chain security, adversarial machine learning, and model governance. AI is often characterized as a system of software, so industry is looking to Software Bills of Materials (SBOMs) as a way to help document and track it during development, operation, maintenance. While traditional SBOMs offer component-level transparency, they fail to capture the dynamic, model-specific risks (e.g. data poisoning, model manipulation) introduced by large language models (LLMs) and other generative AI (GenAI) systems. Similarly, existing transparency artifacts---Model Cards, Datasheets, and System Cards---describe provenance and ethical considerations but seldom provide verifiable, machine-readable evidence tied to threat models. This gap between descriptive documentation and verifiable assurance hinders both security validation and standardized AI risk reporting.

This work introduces the \textbf{AI Risk Scanning (AIRS)} Framework---a threat-model--based, evidence-generating framework for evaluating AI model integrity and behavior. AIRS evolved through three pilot studies that progressively reframed AI assurance from descriptive documentation to verifiable security evidence:
\begin{itemize}[leftmargin=*,nosep]
    \item \textbf{Pilot~A~(Smurf)} developed an \emph{Artificial Intelligence Bill of Materials (AIBOM)}, demonstrating the feasibility of enumerating AI components for transparency.
    \item \textbf{Pilot~B~(OPAL)} validated AIBOM in an operational setting, revealing that transparency alone does not confer assurance without testable evidence and verifiability.
    \item \textbf{Pilot~C~(AIRS)} synthesized these insights into a systematic, machine-verifiable methodology that operationalizes adversarial and supply-chain threats through automated scans and runtime probes.
\end{itemize}

The AIRS Framework aligns its assurance fields with \textbf{MITRE ATLAS} threat categories and extends SBOM standards (SPDX~3.0, CycloneDX~1.6) through structured, evidence-bearing artifacts. Our main contribution is empirical: we outline how AI models can be scanned automatically to produce verifiable, repeatable metrics that reveal contamination, tampering, or unsafe packaging — transforming AI assurance from simple narrative documentation to measurable security evidence.

The contributions in this paper are threefold:
\begin{enumerate}[leftmargin=*,nosep]
    \item \textbf{Conceptual shift:} We expand AI assurance from descriptive reporting to verifiable, evidence-based evaluation aligned with adversarial threat models.
    \item \textbf{Framework design:} We introduce the AIRS Framework, an extensible schema that anchors assurance artifacts to model-level threats and produces machine-readable evidence. Currently, the AIRS Framework is scoped to provide model-level assurances for LLMs, but the framework could be expanded to include other types of models and expand into system-level assurances.
    \item \textbf{Standards integration:} We map AIRS to SPDX~3.0 and CycloneDX~1.6, identifying conceptual overlaps and gaps to guide the evolution of AI-aware SBOMs and ML-BOMs.
\end{enumerate}

% =============================
% 2. PAPER ORGANIZATION
% =============================
\section{Paper Organization}
\label{sec:organization}

The remainder of this paper is organized as follows. 
Section~\ref{sec:related} reviews prior work in AI assurance, 
software supply-chain security, and transparency documentation. 
Section~\ref{sec:method} describes our development methodology, outlining the 
three iterative pilots that evolved from the initial AIBOM schema 
to the operational \emph{AI Risk Scanning (AIRS)} Framework. 
Section~\ref{sec:pilots} presents detailed results and insights from each pilot study—%
Smurf (AIBOM schema design), OPAL (operational validation), and AIRS (automated threat modeling). 
Section~\ref{sec:standards} compares the AIRS Framework to existing SBOM standards, including SPDX~3.0 
and CycloneDX~1.6, and highlights conceptual overlaps and representation gaps. 
Section~\ref{sec:discussion} discusses the broader implications for AI assurance practices, 
while Section~\ref{sec:limits-future} outlines limitations and future research directions. 
Finally, Section~\ref{sec:conclusion} concludes the paper.

% =============================
% 3. RELATED WORK
% =============================
\section{Related Work}
\label{sec:related}
\textbf{GenAI/LLM Threat Models.}
MITRE’s ATLAS~\cite{atlas2023} and the Adversarial ML Threat Matrix~\cite{advmlmatrix2020} adapt ATT\&CK-style tactics and techniques to model-level attacks such as poisoning, unsafe serialization, and backdoors. Application-layer risks like prompt injection, insecure output handling, and supply-chain abuse are emphasized in the OWASP Top 10 for LLMs~\cite{owasp2023} and Google’s Secure AI Framework (SAIF)~\cite{saif2023}, while national guidance (NCSC, CISA)~\cite{ncsc2023,cisa2023} extends secure-by-design principles across training and deployment. NIST’s AI~100-2 taxonomy~\cite{nist1002} and ISO/IEC~23894~\cite{iso23894} provide standardized vocabularies for adversarial ML and risk management. Collectively, these efforts converge on categories spanning training-data compromise, model artifact/supply-chain risks, runtime integration abuse, and misuse at scale, but they seldom attach machine-readable, auditable assurance artifacts.

\textbf{Standards.}
SPDX 3.0 and CycloneDX 1.6 ML-BOM introduce AI entities and component types, enabling exchange of model/dataset metadata~\cite{spdx3spec,cyclonedx15,cyclonedx_mlbom_profile,spdx_ai_profile}. Classical SBOM practice (e.g., National Telecommunications and Information Administration (NTIA) minimum elements, National Institute of Standards and Technology Secure (NIST) Software Development Framework (SSDF), Supply-Chain Levels for Software Artifacts (SSLA)) motivates attestations and hashes~\cite{ntia2021sbommin,nist2022ssdf,slsa1p0} but does not prescribe AI-specific provenance or LLM risk evaluation.

Beyond enumerating fields, we performed a field-level crosswalk between the AIRS Framework and two SBOM standards---SPDX 3.0 and CycloneDX 1.6. We observe strong alignment with SPDX 3.0 on identity and evaluation metadata (e.g., identifiers, licensing, compact benchmark summaries). By contrast, our inspection found little conceptual overlap with CycloneDX 1.6, which remains software-component–centric; most AIRS categories lack native landing zones in CDX as of our review (Oct 2025). Section \ref{sec:standards} summarizes this crosswalk; detailed mappings and gaps appear in Appendix \ref{app:crosswalk}.

% ==== PART 2 of 4 ====
\textbf{Documentation lineage.}
Model Cards~\cite{mitchell2019modelcards}, Datasheets~\cite{gebru2021datasheets}, Data Statements~\cite{bender2018datastatements}, and Dataset Nutrition Labels~\cite{holland2020datanutrition} improve transparency but rarely connect to executable checks or machine-readable evidence. Data Cards~\cite{pushkarna2022datacards} move toward structured metadata but still lack bindings to verifiable security artifacts. Recent System Cards (e.g., GPT-5 System Card~\cite{openai_gpt5_systemcard_2025}, GPT-4o System Card~\cite{openai_gpt4o_systemcard_2024}, ChatGPT Agent System Card~\cite{openai_chatgpt_agent_systemcard_2025}, GPT-4V System Card~\cite{openai_gpt4v_systemcard_2023}) document risks and mitigations at the service level but remain largely descriptive. The AIRS framework draws inspiration from ML documentation practices but prioritizes security-first, evidence-bound fields.

\textbf{Community advances.}
International bodies have called for BOM-like transparency for AI~\cite{bsi_ai_sbom_foodforthoughts}; the Linux Foundation’s SPDX 3.0 workstream proposes AI and Dataset profiles to clarify provenance and obligations~\cite{bennett2024spdx_aibom_guide}. Commercial platforms advertise governance features that resemble BOM capabilities~\cite{noma_ai_governance,procap360}, though contents are not readily examinable. We focus on security-first fields and concrete \emph{assurance artifacts} that bind documentation to verifiable checks.

\textbf{Packaging safety.}
Some LLM packaging methods and formats can be embedded with malicious code that runs when the model is loaded. For example, the common \texttt{pickle} format is not secure against maliciously constructed data ~\cite{jfrog_hf_malicious_models}; stacks that rely on the common machine learning framework \texttt{pytorch} inherit this risk~\cite{python_pickle_docs}. Model hubs promote safer, non-executable formats such as \texttt{safetensors} and employ scanning~\cite{huggingface_safetensors,hf_pickle_scanning}. 

\textbf{Poisoning/contamination \& detection.}
Memorization and extraction risks are well documented~\cite{carlini2021extracting,carlini2023memorization}. Backdoor risks and mitigations are surveyed in~\cite{li2022backdoorSurvey}, with early supply-chain exemplars~\cite{gu2017badnets}; The \emph{BackdoorLLM} benchmark measures the effectiveness of various backdoor attacks ~\cite{backdoorllm2024}. Membership inference attacks~\cite{shokri2017membership,salem2019mlleaks,hu2022membership,ye2022enhanced} motivate contamination probes and privacy-conscious evaluations. We use these underlying techniques for defensive probing and attach results as evidence rather than certifications.

% =============================
% 5. DEVELOPMENT METHODOLOGY
% =============================
\section{Development Methodology}
\label{sec:method}

Our methodology evolved through three consecutive pilot studies that progressively reframed how AI assurance information should be represented---from a descriptive bill of materials to an evidence-driven, threat-model-aligned construct. Each stage refined the schema, artifacts, and verification approach, culminating in the \emph{AIRS} Framework introduced in Pilot~C.

\subsection{Iterative Refinement Across Pilots}
\begin{itemize}[leftmargin=*,nosep]
    \item \emph{Pilot~A~(Smurf)} produced a comprehensive \textbf{Artificial~Intelligence~Bill~of~Materials (AIBOM)}, unifying model, dataset, and operational metadata to support transparency.
    \item \emph{Pilot~B~(OPAL Study)} attempted to validate the AIBOM within an operational GenAI platform and, through practical lessons learned, revealed that enumerating ingredients alone fails to capture cybersecurity risks of AI.
    \item \emph{Pilot~C~(AIRS)} produced a \textbf{threat-model-based, evidence-generating} methodology for LLM security that automatically produces auditable artifacts bound to model-level adversarial behaviors.
\end{itemize}

% =============================
% 6. PILOT STUDIES
% =============================
\section{Pilot Studies}
\label{sec:pilots}

% -----
% Pilot A: SMURF / AIBOM
% -----
\subsection{Pilot A: Smurf - Developing a more Comprehensive Artificial Intelligence Bill of Materials (AIBOM)}
\label{sec:pilotA}

\paragraph*{Motivation and Objectives}
The first pilot and development effort explored whether the ``bill-of-materials'' paradigm used in software assurance could be extended to artificial intelligence systems. Its objective was to design a structured schema---the \emph{AIBOM}---that consolidates technical, operational, and ethical metadata from model and data cards, augmenting them with cyber-relevant fields such as dataset provenance, licensing, quantization, and deployment context.

\paragraph*{Methodology}
The pilot aimed to capture all information required to understand the materials and used to train and evaluate AI models, as well as additional risk fields. It captured a comprehensive schema comprising ten categories: Model~Overview, Model~Details, Training~and~Validation, Testing, Data~Workflow, Performance, Software~and~Hardware~Tools, Deployment, Ethical~Considerations, and Usage~Limits. Appendix~\ref{app:aibom_fields} lists all fields and descriptions.

\paragraph*{Findings}
The AIBOM schema showed the potential for unprecedented supply-chain transparency but revealed three important limitations:
(i)~documentation completeness did not imply security understanding;
(ii)~many entries depended on unverifiable publisher statements; and
(iii)~manual population was unsustainable at scale.
These insights established the need for automated, verifiable security documentation.

\paragraph*{Outcome}
Smurf delivered a comprehensive AIBOM~schema, now serving as a foundational template for AI documentation and a precursor to later, evidence-driven constructs.

% -----
% Pilot B: OPAL / Cyber Robustness Card
% -----
\subsection{Pilot B: OPAL - Validation and Lessons Learned through Production}
\label{sec:pilotB}

\paragraph*{Context}
To evaluate the AIBOM in a production environment, the team applied it within
APL's internal GenAI platform, On-Premise APL Large Language Models (\emph{OPAL}), and concurrently completed our AIBOM on Meta's Llama~3.3~models. The goal was to determine whether AIBOM-style documentation could support security assurance for deployed~LLMs. We describe key lessons learned:

\textbf{(1)~Verifiability must drive scope.}
Many questionnaire items demanded unpublished or unverified information. Fields should focus on verifiable information. Each field could include a verifiability block (\emph{source\_type, URL, confidence, notes}) and accept ``Undisclosed by publisher'' when applicable.

\textbf{(2)~Re-anchor questions to vulnerabilities.}
Questions must map to concrete threat classes: data poisoning~$\rightarrow$~dataset provenance,
prompt injection~$\rightarrow$~publisher mitigations, model theft~$\rightarrow$~licensing and rate-limits, privacy leakage~$\rightarrow$~privacy-preserving training, supply-chain risk~$\rightarrow$~artifact hashes and signing.

\textbf{(3)~Split developer and deployer audiences.}
Developers control training data and internals; deployers manage serving stacks and patches.
Dual tracks prevent dead-ends.

\textbf{(4)~Require machine-readable outputs.}
Replace open-ended prose with structured files such as \texttt{requirements.txt}, container digests, or YAML~blocks capturing serving configurations and evidence metadata.

\textbf{(5)~Add rationale and reviewer actions.}
Each question should include ``Why we ask,'' ``Good answer,'' and ``Reviewer action'' to transform checklists into actionable security controls.

\paragraph*{Cross-Cutting Recommendations}
Add an evidence-confidence rubric, minimal/ideal guidance, red-team hooks, controls mapping to NIST~AI~Risk Management Framework and OWASP~LLM~Top~10, and change-tracking fields (\emph{answer\_date, doc\_version, review\_cycle}).

\paragraph*{Critical Insight}
Through this validation, the team recognized a fundamental limitation: \emph{A bill of materials can list what an AI system contains but not how those ingredients interact to create exploitable vulnerabilities.} Assurance therefore requires a threat-centric approach that produces verifiable,
machine-readable evidence rather than static human-filled templates.

\paragraph*{Outcome}
These lessons catalyzed the development of a new methodological framework -\emph{AI Risk Scanning} -which reframes AI assurance around active threat modeling and automated evidence collection.

% -----
% Pilot C: AIRS / Threat Model
% -----
\subsection{Pilot C: AI Risk Scanning (AIRS) Framework}
\label{sec:pilotC}

\begin{figure*}[htbp]
    \centering
    \includegraphics[width=0.8\linewidth]{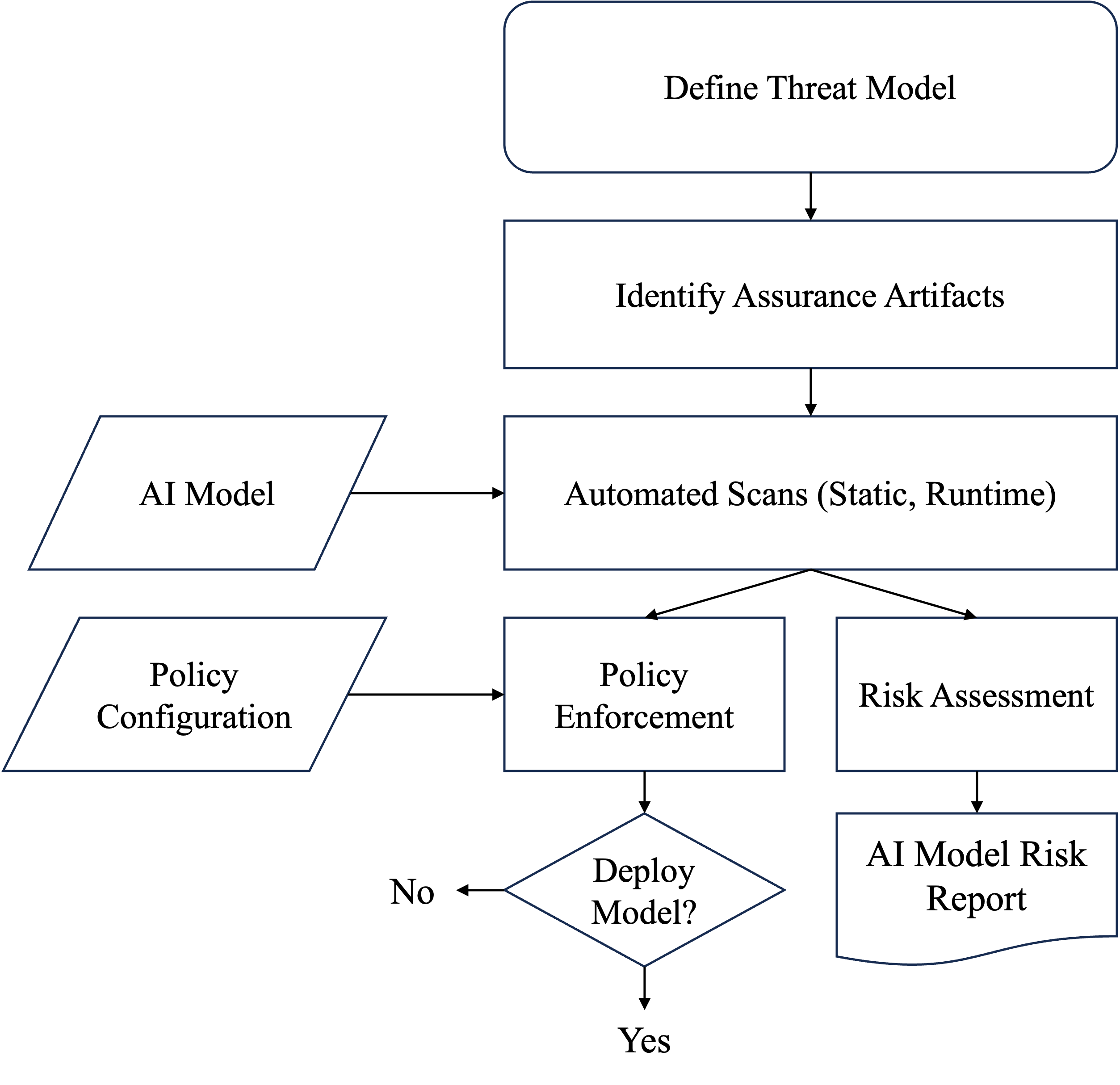}
    \caption{Overview of the \textbf{AI Risk Scanning (AIRS)} Framework. AIRS operationalizes AI assurance through a threat-model-based workflow that produces structured, machine-verifiable evidence. Automated scans enforce user-defined policy configurations and produce AI model risk reports, clarifying risk awareness and informing production monitoring.}
    \label{fig:system-architecture}
\end{figure*}

\paragraph*{Motivation and Conceptual Shift}
Building on the OPAL pilot's findings, Pilot~C replaced enumerative documentation with a \textbf{threat-model-based methodology} that directly tests and evidences model security. The resulting AIRS Framework treats assurance artifacts as first-class outputs of automated scans rather than narrative documentation. Whereas the AIBOM cataloged what an AI system contains, the AIRS Framework verifies AI against known threat models.

\subsubsection{Threat Model and Coverage}
\label{sec:threats}
AIRS scopes its threat coverage to \textbf{model-level adversarial techniques} from MITRE~ATLAS~\cite{atlas2023} that manifest directly within model artifacts - weights, configurations, tokenizers, or serialized packages. Each tactic is operationalized as an evidence-bearing check aligned to AIRS’s categories (Table~\ref{tab:threats}). System-level integrations
(e.g., retrieval connectors, orchestration tools, or deployment posture) are treated as adjacent profiles outside the present scope.

\begin{table*}[!ht]
\centering
\caption{Threats scoped to model-level artifacts, aligned to ATLAS threat types.}
\label{tab:threats}
\begin{tabular}{@{}p{0.03\linewidth}p{0.4\linewidth}p{0.3\linewidth}@{}}
\toprule
\textbf{\#} & \textbf{Threat} & \textbf{Threat Type (ATLAS)} \\
\midrule
1.1 & Poisoned training data & Data Poisoning / Training Data Manipulation \\
1.2 & Publish poisoned or backdoored models & Data Poisoning / Training Data Manipulation \\
1.3 & Erode dataset integrity (misstated versions/splits) & Data Poisoning / Training Data Manipulation \\
\midrule
2.1 & Unsafe deserialization (\texttt{pickle}/\texttt{torch.load}) & AI/ML Supply Chain Compromise \\
2.2 & Non-weight binary loading (bundled \texttt{.so}/\texttt{.dll}) & AI/ML Supply Chain Compromise \\
2.3 & ONNX custom-op injection & AI/ML Supply Chain Compromise \\
2.4 & Masquerading (lookalike model/provider) & AI/ML Supply Chain Compromise \\
\midrule
3.1 & Modify weights (silent edits, NaNs, drift) & Manipulate AI Model \\
3.2 & Modify architecture/config (hidden size, tokenizer merges) & Manipulate AI Model \\
3.3 & Weight steganography (payloads in tensor bit-planes) & Manipulate AI Model \\
3.4 & Adapter/LoRA injection & Manipulate AI Model \\
3.5 & GGUF/metadata prompt injection & Manipulate AI Model \\
\midrule
4.1 & Corrupt shards (damaged or mismatched) & Corrupt AI Model \\
4.2 & Family confusion (obfuscating lineage/variant) & Corrupt AI Model \\
\midrule
5.1 & Backdoor (trigger-activated misbehavior) & Backdoor Insertion / Trigger-based Behaviors \\
5.2 & Tokenizer manipulation as hidden trigger & Backdoor Insertion / Trigger-based Behaviors \\
\midrule
6.1 & Reverse shell on load & Exfiltration and Reverse Shell \\
\midrule
7.1 & Potential for insecure code generation & LLM Response Rendering \\
7.2 & Hallucinated entities (fabricated resources or instructions) & LLM Response Rendering \\
\midrule
8.1 & Data leakage (PII regurgitation) & Data Exposure / Leakage \\
8.2 & Prompt self-replication (embedded prompts) & Data Exposure / Leakage \\
8.3 & Jailbreak susceptibility & Data Exposure / Leakage \\
\bottomrule
\end{tabular}
\end{table*}

\paragraph*{Scope and Coverage}
The AIRS Framework concentrates on threats that are \emph{checkable against model artifacts alone}. Packaging-time issues are mitigated through \textbf{static checks} (integrity validation, serialization safety, tokenizer/graph differences), while runtime risks are captured via \textbf{runtime probes} (sandboxed loads, backdoor/jailbreak sweeps, contamination indicators). 

\textit{To clarify scope, the runtime probes described here operate strictly on model-level artifacts and isolated execution traces; they do not extend to integrated or system-level monitoring.} 

Together, these checks populate AIRS’s machine-readable evidence schema.

\subsubsection*{Framework Design}
At the time of writing, the AIRS Framework organizes 41~fields into five categories with requirement levels: Must~(M), Should~(S), and May~(m). These define which artifacts are mandatory for automation and export while retaining auditable, verifiable intent.

AIRS structures its assurance logic into five evidence-bearing categories:

\begin{enumerate}[leftmargin=*,nosep]
    \item \textbf{Identity \& Release Integrity:} provenance, versioning, and signatures.
    \item \textbf{Packaging \& Serialization Safety:} loader allow-lists and blocked-load logs.
    \item \textbf{Structure \& Adapters:} quantization schemes, adapter integrity, tensor checksums.
    \item \textbf{Runtime Probes:} contamination, backdoor, and jailbreak sweeps with metrics.
    \item \textbf{Evaluation \& Disclosure:} AI test and evaluation benchmark summaries and training-data cutoffs.
\end{enumerate}

\begin{table*}[!t]
\centering
\caption{AIRS fields aligned to ATLAS threats.\ \textit{Legend: Req.\ = M (must), S (should), m (may).}}
\label{tab:crc-fields}
\setlength{\tabcolsep}{4pt}
\begin{tabularx}{\textwidth}{@{}p{0.04\textwidth}p{0.18\textwidth}p{0.25\textwidth}p{0.04\textwidth}X@{}}
\toprule
\textbf{\#} & \textbf{Field (key)} & \textbf{Category} & \textbf{Req.} & \textbf{Description} \\
\midrule
1.1 & model\_name & Identity \& Release Integrity & M & Canonical model name used in packaging/release. \\
1.2 & model\_id & Identity \& Release Integrity & M & Registry path or internal repository slug. \\
1.3 & version\_or\_commit & Identity \& Release Integrity & M & Semantic version or commit hash. \\
1.4 & license & Identity \& Release Integrity & M & License governing model weights. \\
1.5 & hash\_manifest & Identity \& Release Integrity & M & Per-file SHA-256 for shards, configs, and tokenizers. \\
1.6 & signature\_bundle & Identity \& Release Integrity & S & Cryptographic signature bundle for release. \\
1.7 & dir\_merkle & Identity \& Release Integrity & S & Directory-level Merkle root for reproducibility. \\
1.8 & publisher\_evidence & Identity \& Release Integrity & S & Publisher or organization provenance evidence. \\
1.9 & config\_fingerprint & Identity \& Release Integrity & S & Hash of configuration files. \\
1.10 & family\_fingerprint & Identity \& Release Integrity & S & Declared lineage or family fingerprint. \\
\midrule
2.1 & packaging\_policy & Packaging \& Serialization Safety & M & Loader allowlist; block unsafe serializers (e.g., pickle). \\
2.2 & serializer\_scan & Packaging \& Serialization Safety & M & Static scan for disallowed serializers or exec paths. \\
2.3 & guard\_results & Packaging \& Serialization Safety & M & Load-time enforcement outcomes with reasons. \\
2.4 & file\_inventory & Packaging \& Serialization Safety & S & Typed inventory of all package files. \\
2.5 & binary\_inventory & Packaging \& Serialization Safety & S & Inventory of native binaries (\texttt{.so}, \texttt{.dll}). \\
2.6 & allowlist\_policy & Packaging \& Serialization Safety & S & File/extension allowlist policy. \\
2.7 & blocked\_loads\_log & Packaging \& Serialization Safety & m & Evidence of blocked unsafe loads. \\
2.8 & metadata\_scan & Packaging \& Serialization Safety & S & Config metadata diff for hidden templates. \\
2.9 & tokenizer\_fingerprint & Packaging \& Serialization Safety & S & Hashes for tokenizer vocab and merges. \\
2.10 & onnx\_op\_scan & Packaging \& Serialization Safety & S & ONNX/custom-op allowlist and graph lint results. \\
\midrule
3.1 & base\_model & Structure \& Adapters & S & Declared upstream base if finetuned/derived. \\
3.2 & quantization & Structure \& Adapters & S & Quantization scheme and bit-width. \\
3.3 & adapters\_lora & Structure \& Adapters & S & Declared LoRA/adapters and attach points. \\
3.4 & adapter\_inventory & Structure \& Adapters & S & Enumerated attached PEFT/LoRA modules. \\
3.5 & adapter\_hashes & Structure \& Adapters & S & Checksums or signatures for each adapter. \\
3.6 & shape\_consistency\_report & Structure \& Adapters & S & Comparison of config and tensor shapes. \\
3.7 & tensor\_checksums & Structure \& Adapters & S & Per-tensor checksum for drift detection. \\
3.8 & tensor\_stats & Structure \& Adapters & S & Tensor statistics (mean/var/NaN/Inf/outliers). \\
\midrule
4.1 & detector\_method & Runtime Probes & S & Contamination probe method and parameters. \\
4.2 & detector\_outputs & Runtime Probes & S & Probe metrics/curves (ROC/AUC, TPR@5\%FPR). \\
4.3 & backdoor\_probe\_results & Runtime Probes & S & Trigger sweep and attack success rates. \\
4.4 & pii\_probe\_results & Runtime Probes & m & PII exposure and near-duplicate tests. \\
4.5 & jailbreak\_probe\_results & Runtime Probes & m & Jailbreak evaluation results vs.\ baselines. \\
4.6 & prompt\_leak\_probes & Runtime Probes & m & Prompt self-replication/extraction outcomes. \\
4.7 & sanity\_prompts\_diff & Runtime Probes & m & Prompt drift checks for regression testing. \\
4.8 & activation\_probe & Runtime Probes & m & Representation-based anomaly indicators. \\
\midrule
5.1 & benchmark\_summary & Evaluation \& Disclosure & S & Compact summary of evaluation results. \\
5.2 & eval\_datasets & Evaluation \& Disclosure & S & Benchmarks used for evaluation (not training). \\
5.3 & metrics & Evaluation \& Disclosure & S & Reported metrics (accuracy, pass@k, etc.). \\
5.4 & eval\_params & Evaluation \& Disclosure & m & Evaluation configuration (shots, temperature). \\
5.5 & training\_data\_cutoff & Evaluation \& Disclosure & m & Declared training data cutoff date. \\
\bottomrule
\end{tabularx}
\end{table*}

\subsubsection*{Implementation}
To demonstrate a proof-of-concept, an automated AIRS workflow was executed on a 4-bit-quantized \texttt{GPT-OSS-20B} model. Load-time policies restricted deserialization to non-executable formats; per-shard SHA-256 hashes verified integrity; and runtime probes measured behavioral deviations between clean and contaminated inputs.

\paragraph{Policy and Provenance}
A loader policy enforced safe serialization formats (e.g., \texttt{safetensors}, ONNX) and per-shard hash verification. Enforcement outcomes---including policy digests, artifact metadata, and timing---were recorded as signed JSON evidence. Two red-team scenarios validated robustness: (i)~an unsafe \texttt{.pt} artifact blocked on serialization, and (ii)~a single-byte-altered shard blocked on hash mismatch. 
The resulting manifest formed a reproducible provenance trail bound directly to AIRS fields for serialization policy and integrity.

\begin{table}[ht!]
\centering
\caption{Loader guard outcomes.}
\label{tab:guard}
\begin{tabular}{@{}lll@{}}
\toprule
\textbf{Artifact} & \textbf{Serialization} & \textbf{Outcome} \\
\midrule
model-00001-of-00002.safetensors & safetensors & Pass (hash match) \\
unsafe.pt & pickle/pt & Blocked (fail) \\
model\_mutant.safetensors & safetensors & Fail (hash mismatch) \\
\bottomrule
\end{tabular}
\end{table}

\paragraph{Runtime Probes}
To capture dynamic risk, the AIRS Framework implements lightweight runtime probes (see Appendix~\ref{app:perplexity} and Figures~\ref{fig:lp-dist}–\ref{fig:edd-dist}). These probes compute per-sequence log probabilities to detect dataset contamination and poisoning, producing ROC metrics
(AUC, TPR~@~5\%~FPR). Executed deterministically with fixed seeds and sandboxed hardware, each run emits per-item logs and summary metrics. Outputs are serialized as portable, auditable artifacts and treated as triage signals rather than certification results.

\begin{table}[ht!]
\centering
\caption{Runtime probe micro-pilot (N=30).}
\label{tab:detectors}
\begin{tabular}{@{}lll@{}}
\toprule
\textbf{Probe} & \textbf{AUC} & \textbf{TPR @ 5\% FPR} \\
\midrule
Log-prob (Control) & 0.512 & 0.267 \\
Log-prob (Contaminated) & 0.646 & 0.500 \\
\bottomrule
\end{tabular}
\end{table}

\subsection{Risk Scoring as a Design Consideration}
While AIRS provides structured, evidence-bound categories, risk scoring remains profile-dependent. Single scalar scores are often too coarse; we therefore propose \emph{profiled interpretations} that weight categories differently. Designing and calibrating such profiles is future work; AIRS is structured to support it.

\paragraph*{Outcome}
The AIRS Framework demonstrated that automated, evidence-centric scans can produce verifiable artifacts directly mapped to model-level threats. Collectively, the three pilots mark a progression---from \emph{enumerating AI components for transparency}~(AIBOM), to \emph{validating and revealing the limits of BOM-based assurance for AI}, to \emph{providing measurable, threat-oriented coverage through AIRS’s automated evidence generation}.

% =============================
% 7. AIBOM AND STANDARDS
% =============================
\section{AIRS and Standards}
\label{sec:standards}
\subsection{Context and Goal}
As SBOM practices mature in AI procurement and assurance workflows, it is natural to ask where the AIRS Framework sits relative to widely used specifications, e.g., SPDX 3.0 and CycloneDX 1.6. Our intent is not to create a competing framework, but to show how AIRS can \emph{extend} upon them. Concretely, we (i) surface the conceptual overlaps that enable compatibility between AIRS fields and existing SBOM constructs, and (ii) make explicit the places where AI-security evidence (e.g., loader policies, runtime probe outputs) is not yet modeled as explicitly requested evidence within the SBOM structure and therefore deserves careful standardization. In practice, we treat SPDX 3.0 as the primary near-term export target and report CycloneDX 1.6 results for completeness, noting the minimal overlap observed and therefore opportunity for collaboration.

\subsection{AIRS \texorpdfstring{$\leftrightarrow$}{↔} SPDX 3.0 and CycloneDX 1.6: Empirical Crosswalk}
\label{sec:mapping}
We conducted a field-by-field crosswalk between AIRS entries and elements available in 
\textbf{SPDX 3.0} and \textbf{CycloneDX 1.6}. Two clear patterns emerged. 

\textbf{First}, \emph{SPDX 3.0 exhibits strong conceptual overlap} with AIRS categories centered on \emph{Identity \& Release Integrity} and \emph{Evaluation \& Disclosure}. Examples include model identifiers, licensing, and compact benchmark summaries, which map naturally onto SPDX constructs such as package identity, license expressions, and annotation metadata. This overlap suggests SPDX 3.0 is a practical near-term anchor for AIRS export. 

\textbf{Second}, \emph{CycloneDX 1.6 shows minimal overlap}. Our inspection (as of October 2025) found that CDX remains primarily software-component–centric, with most LM threat categories lacking explicit fields. Apart from basic identity elements, fields tied to packaging and serialization behavior, structural modifications (e.g., quantization, adapters, tokenizer fingerprints), and runtime probes had no clear conceptual counterparts. 

Across both standards, three AIRS categories remain \emph{unique and security-critical} for model assurance today: 
\emph{Packaging \& Serialization Safety} (loader/serializer policies and unsafe-path findings), 
\emph{Structure \& Adapters} (quantization, adapters, tokenizer integrity), and 
\emph{Runtime Probes} (portable summaries of probe configuration and outcomes). 
Current SBOM specifications do not yet include these as first-class elements. 
Section~\ref{sec:future-integration} outlines candidate directions for encoding such evidence in existing SBOM ecosystems, while Appendix~\ref{app:crosswalk} provides the detailed field-level comparison.

\subsection{Design Intent for SBOM Extension and Future Validation}
\label{sec:future-integration}
Our intent is to extend existing SBOM and ML-BOM frameworks rather than replace or parallel them. AIRS is conceived as a set of complementary descriptors that enrich current SBOM formats with AI-specific assurance artifacts. We envision three areas of extension: (i) descriptors for serializer and loader policies, including unsafe-path findings, integrated under existing packaging and dependency concepts; (ii) structured, auditable encodings for adapter, quantization, and tokenizer structures, aligned with model-component representations already present in ML-BOMs; and (iii) typed \emph{runtime evidence} records capturing scan configurations, datasets or prompts, and corresponding summary statistics. 

% =============================
% 8. DISCUSSION
% =============================
\section{Discussion}
\label{sec:discussion}
Security documentation for AI must evolve beyond static, narrative Model and Data Cards. Even a comprehensive bill of materials cannot by itself detect or characterize AI threats. Instead, assurance should move toward \textbf{automated, threat-informed assessments} that connect risks to verifiable evidence. The AIRS framework embodies this shift by anchoring its fields to model-level, ATLAS-aligned threats, emphasizing assurance artifacts, and incorporating mechanisms for evaluating dynamic runtime risks. This approach reflects two key design choices: we prioritize \emph{verifiable evidence} over unverifiable disclosures and capture dynamic risks through runtime probes whose results become portable assurance artifacts.

This perspective reflects lessons learned across three progressive pilots. \textbf{Pilot~A~(Smurf)} demonstrated that enumerating model and dataset elements improved transparency but did not translate into measurable security confidence. \textbf{Pilot~B~(OPAL)} showed that verifiability-rather than completeness-must drive documentation scope, grounding each field in observable or testable evidence linked to concrete threat classes. \textbf{Pilot~C~(AIRS)} advanced these insights into a threat-model-based methodology that treats assurance as an active, evidence-producing process rather than static recordkeeping. The resulting scope of AIRS remains intentionally conservative, focusing on what can be verified from model artifacts and bounded runtime probes, while deferring orchestration and tooling to adjacent profiles for tractability and auditability.

In positioning AIRS within existing standards, this evolution clarifies why a conventional SBOM alone is insufficient for AI assurance. The limited conceptual overlap observed with \textbf{CycloneDX~1.6} underscores the need for richer, security-first descriptors, whereas \textbf{SPDX~3.0} provides a more natural near-term export target for AIRS’s structured, evidence-bound fields. AIRS thus complements and extends SBOM practices by adding the verifiable, threat-linked evidence required for trustworthy AI provenance.

% ==== CONTINUATION of PART 4 ====
% =============================
% 9. LIMITATIONS AND FUTURE WORK
% =============================
\section{Limitations and Future Work}
\label{sec:limits-future}
We summarize limitations and next steps along six axes.

\subsection*{(1) Scope}
\textbf{Current focus (model-level).} Our implementation of AIRS, the threat model, and evidence bindings are scoped to model artifacts and bounded runtime probes. We do not cover tool use, connectors, RAG stores, AI agents, or deployment posture beyond minimal hosting context. This improves tractability and verifiability but under-approximates end-to-end risk.

\subsection*{(2) Threat Model}
\textbf{ATLAS-derived and general.} Our threat model refines MITRE ATLAS tactics to model-level manifestations. This lens is well suited for artifact-centric checks (serialization, integrity, tokenizer differences, backdoor triggers), but it omits application-layer abuses (prompt injection via tools, insecure output handling) and organization-specific threats. As formats, loading paths, and deployment practices evolve---and as usage patterns shift (local/offline vs.\ external API)---We will leverage insights and lessons learned to expand the threat model and add \emph{system-level profiles} that incorporate application-layer catalogs with evidence hooks (e.g., tool-call allowlists, taint tracking, output sanitizers), while remaining auditable.

\subsection*{(3) Limitations of Runtime Probes}
\textbf{Method immaturity.} Our contamination probes (Figures~\ref{fig:lp-dist}–\ref{fig:edd-dist}) are informative but not definitive. Small-sample studies and overlapping distributions highlight the need for stronger methods and careful calibration. Probes are for triage and regression, not acceptance testing.

\subsection*{(4) Must/Should/May and Cyber Security Risk Interpretation}
\textbf{Profiles over single scores.} Requirement levels are helpful for linting but insufficient for decisions, and a single global “risk score” is misleading across use cases. We advocate \emph{profiled} interpretations (task- and environment-specific bundles) with transparent weighting and sensitivity analysis, enabling explanations of why two models rank differently under different profiles.

\subsection*{(5) Modality Scope---LLMs Only}
This work targets text-only LLMs, and the AIRS Framework categories, validators, and runtime probes are designed around LLM artifacts and risks. While some evidence types transfer (e.g., integrity, serializer/custom-op scans), runtime probes and packaging are calibrated for text. Extending to other modalities will require an expanded threat model with additional fields while maintaining similar principles of verifiability and measurement of dynamic risks connected with threat models.

\subsection*{(6) Standards synthesis} 
The crosswalk provides practical guidance without assuming conversion: treat AIRS identity and evaluation fields as conceptually aligned with existing SBOM constructs, and preserve AIRS-unique material---packaging/serialization safety, structure/adapters, and runtime probes---as \emph{assurance artifacts}. This framing clarifies where current specifications already speak the language of provenance and integrity, and where AI-security semantics remain to be standardized and validated.

\subsection*{(7) Additional limitations}
Our pilot studies are limited in breadth (one automated single-model pipeline, one internal multi-LLM environment, one platform). We have not yet demonstrated AIRS-to-standard exports at scale, nor longitudinal drift detection using tensor-level integrity in production. Some disclosures (e.g., license, lineage) may rely on publisher assertions; our stance is to call out unverifiable claims and prefer signatures and hash-based evidence where possible.

% =============================
% 10. CONCLUSION
% =============================
\section{Conclusion}
\label{sec:conclusion}

Through a three-pilot progression-from \textbf{AIBOM} to \textbf{OPAL} to \textbf{AIRS}-this work reframes AI assurance as a verifiable, threat-informed discipline rather than a documentation exercise. AIRS demonstrates that automated, evidence-centric scanning can directly generate artifacts aligned to MITRE~ATLAS threat categories, enabling machine-verifiable assurance of model integrity, packaging safety, and runtime behavior. In doing so, it complements-but does not replace-existing SBOM standards, extending them with AI-specific security semantics absent in current specifications. While the current AIRS Framework is scoped to provide model-level assurances for LLMs, the framework can expand to include other types of models and scan for system-level assurances.

Our analysis highlights that \textbf{SPDX~3.0} offers strong structural alignment for AI assurance exports, while \textbf{CycloneDX~1.6} remains limited to software-component views. The unique AIRS categories-\emph{Packaging \& Serialization Safety}, \emph{Structure \& Adapters}, and \emph{Runtime Probes}-represent essential extensions for a trustworthy AI supply chain. Future work will focus on integrating these elements into SBOM ecosystems, expanding threat coverage to application-layer risks, and refining runtime probe methodologies for broader modality and deployment contexts.

In summary, the AIRS Framework provides a foundation for \textbf{evidence-driven AI security assurance}-one that connects model-level threats, verifiable artifacts, and standardization pathways toward transparent, trustworthy, and auditable AI systems.

\section*{Acknowledgments}
We acknowledge JHU/APL for financial support of this research, including supplemental funding that enabled broader stakeholder engagement and greater technical advancement. We also thank our senior technical advisors for feedback and guidance that helped refine our approach and anticipate potential blind spots.

\bibliographystyle{IEEEtran}
\bibliography{refs}

% \clearpage
\appendices
\clearpage
\onecolumn
\section{AIRS \texorpdfstring{$\leftrightarrow$}{↔} SPDX 3.0 and CycloneDX 1.6 Crosswalk}
\label{app:crosswalk}

\vspace{1em}
\newcommand{\yes}{\colorbox{green!20}{\textbf{Y}}}
\newcommand{\no}{\colorbox{red!15}{\textbf{N}}}
\newcommand{\tbd}{\colorbox{gray!20}{\textbf{TBD}}}
\newcommand{\ovl}[1]{\colorbox{green!10}{#1}}

\setlength{\LTpre}{0pt}
\setlength{\LTpost}{6pt}
\renewcommand{\arraystretch}{1.15}

\begin{center}\scriptsize
\begin{longtable}{@{}p{0.05\textwidth}p{0.14\textwidth}p{0.20\textwidth}p{0.07\textwidth}p{0.07\textwidth}p{0.33\textwidth}@{}}
\caption{\centering AIRS: model-level fields with SPDX 3.0 and CycloneDX 1.6 conceptual coverage.\\
\centering\textit{Coverage: \yes\ = conceptual overlap (field name also highlighted), \no\ = no clear counterpart.}}\\
\toprule
\textbf{\#} & \textbf{Field (key)} & \textbf{Category} & \textbf{SPDX 3.0} & \textbf{CDX 1.6} & \textbf{Description (model-level, ATLAS-relevant)} \\
\midrule
\endfirsthead
\multicolumn{6}{l}{\footnotesize\emph{Table (continued)}}\\
\toprule
\textbf{\#} & \textbf{Field (key)} & \textbf{Category} & \textbf{SPDX 3.0} & \textbf{CDX 1.6} & \textbf{Description} \\
\midrule
\endhead
\midrule
\multicolumn{6}{r}{\footnotesize\emph{Continued on next page}}\\
\endfoot
\bottomrule
\endlastfoot

1.1 & \ovl{model\_name} & Identity \& Release Integrity & \yes & \yes & Canonical model name used in packaging/release. \\
1.2 & \ovl{model\_id} & Identity \& Release Integrity & \yes & \yes & Registry path / repo slug or internal ID. \\
1.3 & \ovl{version\_or\_commit} & Identity \& Release Integrity & \yes & \yes & SemVer or commit hash of released artifacts. \\
1.4 & \ovl{license} & Identity \& Release Integrity & \yes & \yes & License for the model weights. \\
1.5 & \ovl{hash\_manifest} & Identity \& Release Integrity & \no & \yes & Per-file SHA256 (all shards/config/tokenizer). \\
1.6 & signature\_bundle & Identity \& Release Integrity & \no & \no & Release/signing bundle for artifacts. \\
1.7 & dir\_merkle & Identity \& Release Integrity & \no & \no & Directory-level Merkle root across the model directory. \\
1.8 & publisher\_evidence & Identity \& Release Integrity & \no & \no & Publisher/org evidence for provenance. \\
1.9 & config\_fingerprint & Identity \& Release Integrity & \no & \no & Hash of \texttt{config.json}/\texttt{model\_index.json}. \\
1.10 & family\_fingerprint & Identity \& Release Integrity & \no & \no & Declared lineage/family fingerprint (non-deceptive). \\
2.1 & packaging\_policy & Packaging \& Serialization Safety & \no & \no & Loader allowlist (\texttt{safetensors}/ONNX); block pickle/\texttt{.pt}. \\
2.2 & serializer\_scan & Packaging \& Serialization Safety & \no & \no & Static scan of files for disallowed serializers/exec paths. \\
2.3 & guard\_results & Packaging \& Serialization Safety & \no & \no & Load-time enforcement outcomes (pass/fail + reason). \\
2.4 & \ovl{file\_inventory} & Packaging \& Serialization Safety & \no & \yes & Typed inventory of every file in package. \\
2.5 & \ovl{binary\_inventory} & Packaging \& Serialization Safety & \no & \yes & Inventory of native binaries (\texttt{.so}/\texttt{.dll}) if present. \\
3.1 & \ovl{base\_model} & Structure \& Adapters & \yes & \no & Declared upstream base (if finetuned/derived). \\
3.2 & quantization & Structure \& Adapters & \no & \no & Quantization scheme/bit-width for packaged model. \\
3.3 & adapters\_lora & Structure \& Adapters & \no & \no & Declared adapters and attach points (if any). \\
3.4 & adapter\_inventory & Structure \& Adapters & \no & \no & Enumerated attached PEFT/LoRA modules. \\
4.1 & detector\_method & Runtime Probes & \no & \no & Contamination probe method (logprob) + params. \\
4.2 & detector\_outputs & Runtime Probes & \no & \no & Probe metrics/curves (ROC/AUC, TPR @ 5 \% FPR). \\
4.3 & backdoor\_probe\_results & Runtime Probes & \no & \no & Trigger sweep + attack success rates; neg. controls. \\
5.1 & \ovl{benchmark\_summary} & Evaluation \& Disclosure & \yes & \no & Compact summary of reported evaluation results. \\
5.2 & \ovl{eval\_datasets} & Evaluation \& Disclosure & \yes & \no & Named benchmarks used for evaluation (not training). \\
5.3 & \ovl{metrics} & Evaluation \& Disclosure & \yes & \no & Reported metrics (accuracy, pass@k, etc.). \\
5.5 & \ovl{training\_data\_cutoff} & Evaluation \& Disclosure & \yes & \no & Declared training data cutoff date (disclosure only). \\
\end{longtable}
\end{center}

\clearpage
\onecolumn
\section{Artificial Intelligence Bill of Materials (AIBOM) Field Schema}
\label{app:aibom_fields}

This appendix summarizes the complete field schema developed in
Pilot~Study~A~(SMURF). Only field names and descriptions are listed;
values are omitted. The schema organizes metadata across ten categories
spanning the model lifecycle.

\setlength{\LTpre}{0pt}
\setlength{\LTpost}{0pt}
\renewcommand{\arraystretch}{1.15}

\begin{longtable}{@{}p{0.35\textwidth}p{0.50\textwidth}@{}}
\caption{Artificial Intelligence Bill of Materials (AIBOM) Field Schema}\\
\toprule
\textbf{Field} & \textbf{Description} \\
\midrule
\endfirsthead
\multicolumn{2}{l}{\footnotesize\emph{Table (continued)}}\\
\toprule
\textbf{Field} & \textbf{Description} \\
\midrule
\endhead
\bottomrule
\endfoot

% -------------------------------
\multicolumn{2}{@{}l}{\textbf{A1. Model Overview}}\\
\midrule
Model Usage & Describes the model’s purpose and intended function. \\
Developer Contact & Contact information for model owner or organization. \\
Model Creation and Release Dates & Dates of model creation and public release. \\
Licenses & Licenses governing model use and redistribution. \\
Hosting Requirements & Hardware requirements for deployment (GPU, RAM, etc.). \\
Software Hosting Requirements & Software dependencies and version constraints. \\
Installation and Use Instructions & Step-by-step instructions or API access notes. \\
Input Format & Expected data format and representative example. \\
Output Format & Structure and examples of model outputs. \\
Operational Graphic & Diagram of model workflow (input–output path). \\
\addlinespace[1ex]

% -------------------------------
\multicolumn{2}{@{}l}{\textbf{A2. Model Details}}\\
\midrule
Architecture Diagram & Graphic illustrating model architecture. \\
Customizations & Description of fine-tuning or architectural changes. \\
Parameter Count & Total number of model parameters. \\
Layer Composition & Enumerated layers and functional components. \\
Model Size & Storage footprint of model artifacts. \\
References & Publications or documentation referencing the model. \\
\addlinespace[1ex]

% -------------------------------
\multicolumn{2}{@{}l}{\textbf{A3. Training and Validation}}\\
\midrule
Training Data Sources & Datasets used for training and validation. \\
Data Format Examples & Illustrative instance for each dataset. \\
Data Class Distributions & Distribution of classes or labels. \\
Dataset Sizes & Data volume in MB/GB/TB. \\
Sampling Methods & Procedures for dataset selection and sampling. \\
Data Relationships & Relationships among data elements. \\
Data References & Citations and source links for datasets. \\
\addlinespace[1ex]

% -------------------------------
\multicolumn{2}{@{}l}{\textbf{A4. Testing}}\\
\midrule
Testing Datasets & Benchmarks used for evaluation or testing. \\
Test Data Format & Structure of test instances. \\
Test Class Distributions & Class proportions in test datasets. \\
Test Dataset Sizes & Size and count of test datasets. \\
Sampling Methods & Approach for generating or selecting test samples. \\
Known Errors and Biases & Documented limitations or error sources. \\
Underrepresented Classes & Categories with limited representation. \\
Synthetic or Augmented Data & Created datasets for addressing outliers. \\
Dataset Relationships & Cross-dataset or intra-dataset dependencies. \\
Testing References & References to dataset cards or publications. \\
\addlinespace[1ex]

% -------------------------------
\multicolumn{2}{@{}l}{\textbf{A5. Data Workflow}}\\
\midrule
Data Ingestion & Steps for converting raw data to model-ready inputs. \\
Data Storage & Storage location or persistence mechanism. \\
Metadata Enrichment & Added annotations or transformations. \\
Computation Engines & Analytic or processing engines used. \\
Pipeline Services & Tools for orchestration and observability. \\
Data Governance & Compliance or policy-enforcement mechanisms. \\
\addlinespace[1ex]

% -------------------------------
\multicolumn{2}{@{}l}{\textbf{A6. Performance}}\\
\midrule
Evaluation Metrics & Metrics for validity and efficiency (accuracy, BLEU, etc.). \\
Testing Parameters & Evaluation configurations (data slices, shots, etc.). \\
Performance Results & Measured accuracy or latency results. \\
Interpretation & Human-readable explanation of results. \\
Trade-offs & Performance versus resource trade-offs. \\
Monitoring Plans & Post-deployment performance tracking. \\
Environmental Impact & Estimated energy use or emissions. \\
Comparative Analysis & Comparison to other models or benchmarks. \\
\addlinespace[1ex]

% -------------------------------
\multicolumn{2}{@{}l}{\textbf{A7. Software and Hardware Tools}}\\
\midrule
Software Dependencies & Libraries or frameworks used to build/test the model. \\
MLOps Pipeline Tools & Tools for training, inferencing, or analytics. \\
Hardware Resources & Compute hardware used in development (GPU, CPU, TPU). \\
\addlinespace[1ex]

% -------------------------------
\multicolumn{2}{@{}l}{\textbf{A8. Deployment}}\\
\midrule
Deployment Platform & Platform or environment used for model hosting. \\
Monitoring Mechanisms & Tools for monitoring model performance post-deployment. \\
Runtime Hardware & Required runtime hardware resources. \\
\addlinespace[1ex]

% -------------------------------
\multicolumn{2}{@{}l}{\textbf{A9. Ethical Considerations}}\\
\midrule
Potential Biases & Known systematic or dataset-induced biases. \\
Fairness Concerns & Potential inequities or disparate impacts. \\
Other Trustworthiness Issues & Transparency, privacy, or compliance concerns. \\
Mitigations & Steps taken to reduce ethical or fairness risks. \\
\addlinespace[1ex]

% -------------------------------
\multicolumn{2}{@{}l}{\textbf{A10. Usage Limits}}\\
\midrule
Context Constraints & Domains where performance may degrade. \\
Harmful Uses & Scenarios where outputs may be unsafe or biased. \\
Other Risks & Additional operational or societal risks. \\
Mitigations & Safeguards or controls to prevent misuse. \\
\end{longtable}

\twocolumn

% =============================
% Appendix B: LLM Runtime Probes
% =============================% =============================
% Appendix B: LLM Runtime Probes
% =============================
\section{LLM Runtime Probes}
\label{app:perplexity}

This appendix summarizes two lightweight probes - log-probability and edit-distance analyses - used to detect possible dataset contamination or memorization. Both generate auditable, model-only evidence suitable for inclusion as AIRS runtime artifacts.

\subsection{Perplexity / Log-Probability Probe}
\textit{Objective.} Estimate contamination by comparing next-token log-probabilities on items known to be \emph{in-training} (true positives, TP) versus \emph{out-of-training} (true negatives, TN). A strong separation implies contamination risk; overlap suggests low signal.

\textit{Setup.} Four candidate datasets were evaluated:
\begin{itemize}[leftmargin=*,nosep]
  \item \texttt{allenai/ultrafeedback\_binarized\_cleaned}
  \item \texttt{allenai/tulu-v2-sft-mixture}
  \item \texttt{allenai/tulu-3-sft-olmo-2-mixture-0225}
  \item \texttt{allenai/OLMo-mix-1124}
\end{itemize}
For \texttt{OLMo-7B-Instruct}, the first two were treated as TP and the latter two as TN sets.

\textit{Method.} Each item’s next-token log-probability was computed at a fixed interior position using a consistent tokenizer. TP and TN distributions were compared via ROC/AUC and recall at 5 \% FPR; one-way ANOVA served as a descriptive check.

\begin{figure}[h]
  \centering
  \includegraphics[width=\linewidth]{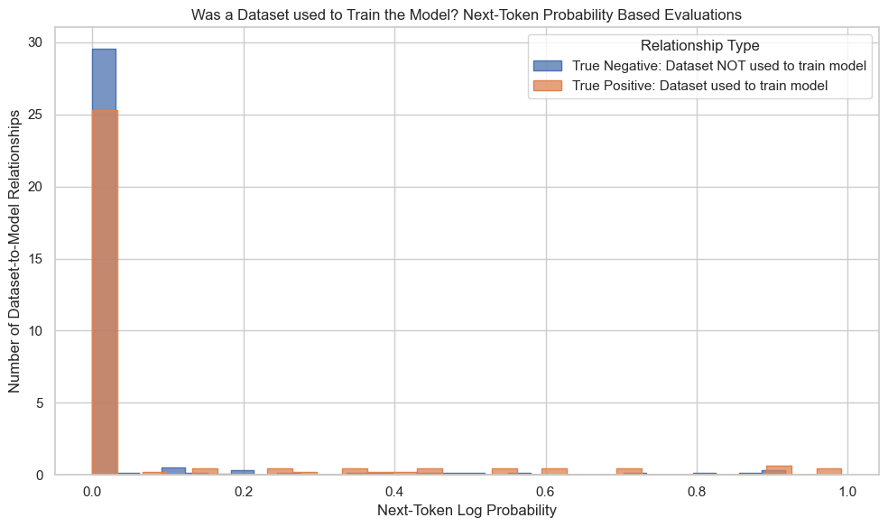}
  \caption{Next-token log-probability distributions for TP vs.\ TN. Overlap motivates richer probes.}
  \label{fig:lp-dist}
\end{figure}

\begin{figure}[h]
  \centering
  \includegraphics[width=\linewidth]{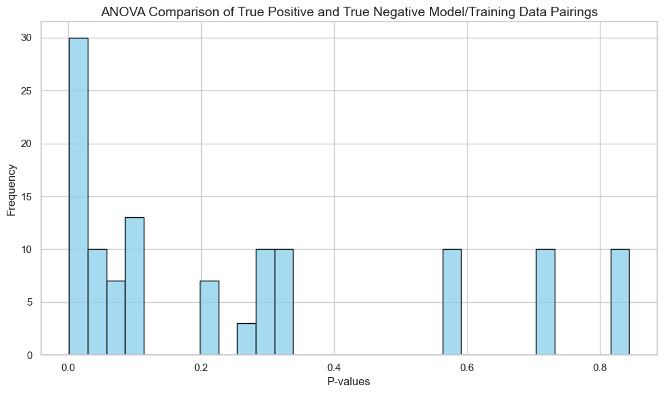}
  \caption{$p$-value distribution from repeated one-way ANOVA. Mixed significance highlights weak separability.}
  \label{fig:lp-anova}
\end{figure}

\textit{Results.}  
Substantial overlap between TP and TN distributions (Fig.~\ref{fig:lp-dist}) and mixed ANOVA significance (Fig.~\ref{fig:lp-anova}) indicate that single-position log-prob probes are weak contamination indicators but remain useful for triage and regression testing.

\textit{Metric interpretation.}  
AUC denotes the probability that a randomly chosen contaminated item ranks higher than a clean one ($\mathrm{AUC}=0.5$ = random).  
TPR @ 5 \% FPR reflects recall at a low false-positive rate---useful for screening, not certification.

\subsection{Edit-Distance Distribution (Exploratory)}
\label{app:edd}

\textit{Objective.} Evaluate \emph{edit-distance distribution (EDD)} as a complementary probe for memorization. EDD measures normalized Levenshtein distance between multiple model responses to the same prompt under deterministic and stochastic decoding.

\textit{Method.} Following~\cite{carlini2021extracting,carlini2023memorization}, a baseline output was generated at temperature 0, then $n$ stochastic samples at temperature $\tau\!>\!0$. Edit distances between stochastic and baseline outputs were compared for contaminated vs.\ clean prompts.

\begin{figure}[h]
  \centering
  \includegraphics[width=\linewidth]{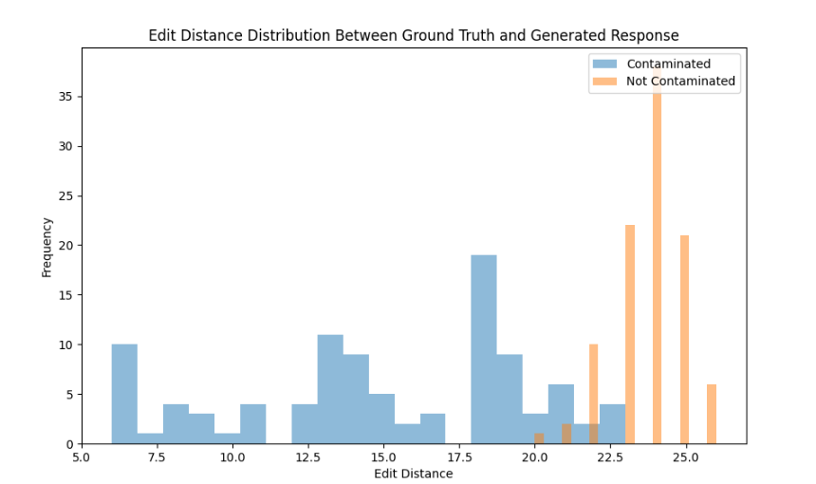}
  \caption{Edit-distance distributions. Contaminated prompts (red) yield more similar outputs than uncontaminated prompts (blue).}
  \label{fig:edd-dist}
\end{figure}

\textit{Findings.} Contaminated samples show lower edit distances (higher similarity) than clean ones (Fig.~\ref{fig:edd-dist}), indicating memorization tendencies. The study was small-scale and should be viewed as illustrative rather than confirmatory.

\textit{Limitations.} EDD requires stochastic sampling and threshold selection not yet standardized; broader evaluation is needed. Nonetheless, its complementary signal supports continued exploration alongside log-prob probes and other membership inference methods for future AIRS runtime checks.

% ==== END OF DOCUMENT ====
\end{document}